\documentclass[aps,prl,superscriptaddress,showpacs,floatfix,twocolumn]{revtex4}
\usepackage{times}
\usepackage{graphicx}% Include figure files
\usepackage{amsmath}
\usepackage{amssymb}
\usepackage{subfigure}
\usepackage{color}

\newcommand{\mP}{{\bf P}}
\newcommand{\mT}{{\bf T}}
\newcommand{\mK}{{\bf K}}
\newcommand{\mM}{{\bf M}}

\begin{document}

\title{Emergent topological mirror insulator in $t_{2g}$-orbital systems}

\author{Yuan-Yen Tai}
\affiliation{Texas Center for Superconductivity \& Department of Physics, University of Houston, Houston, Texas 77004, USA}
\affiliation{Theoretical Division, Los Alamos National Laboratory, Los Alamos, New Mexico 87545, USA}
\author{C.-C. Joseph Wang}
\affiliation{Theoretical Division, Los Alamos National Laboratory, Los Alamos, New Mexico 87545, USA}
\affiliation{Center for Nonlinear Studies, Los Alamos National Laboratory, Los Alamos, New Mexico 87545, USA}

\author{Matthias J. Graf}
\affiliation{Theoretical Division, Los Alamos National Laboratory, Los Alamos, New Mexico 87545, USA}

\author{Jian-Xin Zhu}
\affiliation{Theoretical Division, Los Alamos National Laboratory, Los Alamos, New Mexico 87545, USA}
\affiliation{Center for Integrated Nanotechnologies, Los Alamos National Laboratory, Los Alamos, New Mexico 87545, USA}

\author{C. S. Ting}
\affiliation{Texas Center for Superconductivity \& Department of Physics, University of Houston, Houston, Texas 77004, USA}

\date{\today}

\begin{abstract}
Motivated by the itinerant band structure of high-$T_c$ iron pnictides, which exhibit four Dirac cones in the bulk, we demonstrate the prospect of pnictides with transition elements to be topological insulators in two dimensions.
In this report, we explore interaction-induced topological phases, in contrast to the spin-orbit-coupling interaction, as the crucial mechanism for tuning Dirac metals into Z$_{2}$-topological insulators protected by time reversal and mirror symmetries.
We find spontaneous orbital currents generated through nearest-neighbor inter-orbital Coulomb interaction in the $t_{2g}$ manifold of the $d$ orbitals.
When spin degrees of freedom are incorporated, spontaneous orbital currents lead to two stable topological phases of the ground state.
The first topological insulator is an anomalous orbital Hall phase, characterized by an even Chern number, while the second topological insulator is realized by protected mirror symmetries with a Z$_2$ index.
\end{abstract}
%\pacs{}
\pacs{71.10.Fd,71.10.Pm,73.20.-r}

%-----------------------------------------------------------------------------------------
\maketitle
\paragraph{Introduction.$-$}
Topological insulators (TIs) are typically characterized by the band topology of their electronic wave function in the bulk, which is connected to protected edge or surface states.
In fact, the integer quantum Hall insulator is the first known TI,  which Thouless  and collaborators \cite{DJThouless} characterized by the topological Chern number under the condition of broken time-reversal (TR) symmetry due to an external magnetic field. In the absence of an external magnetic field, the analogs of the quantum Hall effect and topological Chern number were discussed by Haldane \cite{Haldane} for the honey-comb lattice with spontaneous internal magnetic fields between two different sub-lattices, and by Volovik \cite{Volovik1988} for charged and neutral superfluids with $^3$He-like order parameters.
By promoting Haldane's model to a spinful version that respects the TR symmetry in the presence of strong spin-orbit coupling (SOC),  
the concept of the quantum-spin Hall insulator was proposed~\cite{Kane}, which is characterized by a nontrivial $Z_{2}$ topological invariance.
The manifestation of TIs with $Z_{2}$ symmetry is accompanied by the opening of a gap in Dirac semi-metals due to the SOC interaction, and the emergence of gapless,  symmetry-protected edge (or surface) states in two (or three) dimensions.

Ever since the discovery of TIs, new types and realizations have been extended to new materials \cite{Knig,Bernevig,Hsieh,Zhang,KAI,WEN,Teo, Klintenberg,DZERO,Fu3,Alexandrov2013, Hasan,Qi2011,Hohenadler2013}.
In this letter, we show a very different realization of the topological mirror insulator (TMI)~\cite{Hsieh1,Fiete} in the $t_{2g}$ bands of two-dimensional (2D) insulators with pnictide-like band structure.
In contrast to the insulating gap generated by the SOC models for graphene or bismuth, \cite{Hasan, Ando}
or interaction-driven topological insulators in the presence of strong SOC, 
we propose a new route to realizing non-trivial, emergent topological phases within the $t_{2g}$ low-energy manifold in transition element materials \cite{Khomskii} with the insulating gap opened by purely correlated electron interactions~\cite{Raghu},
see also Refs.~\cite{Qi2011, Hohenadler2013} and references therein.
Our discussion is based on a realistic, minimal quasi-2D model~\cite{YTai,HChen,footnote2}, which proved successful in reproducing the electronic structure and phase diagram of the 122 iron pnictides.
In this two orbital model, we consider onsite intra-orbital and nearest-neighbor (NN) inter-orbital Coulomb interaction treated within mean-field theory.
With reasonable hopping and Coulomb parameters, we find spontaneous orbital currents in the ground state. These orbital currents generate non-trivial topological phases with two pairs of Dirac cones appearing as edge states.
We further show that a non-trivial mirror-$Z_2$ phase can be identified for the spinful Hamiltonian.
This novel phase involves mirror reflection symmetries in 2D and is robust against weak TR breaking perturbations.
To be more specific, this phase can only be destroyed by perturbations that break the mirror symmetry in the Brillouin zone (BZ).
Hence the presented scenario is markedly different from previous TIs, which are protected by TR symmetry and exhibit an odd number of Dirac cones.

\begin{figure}
\includegraphics[scale=0.35]{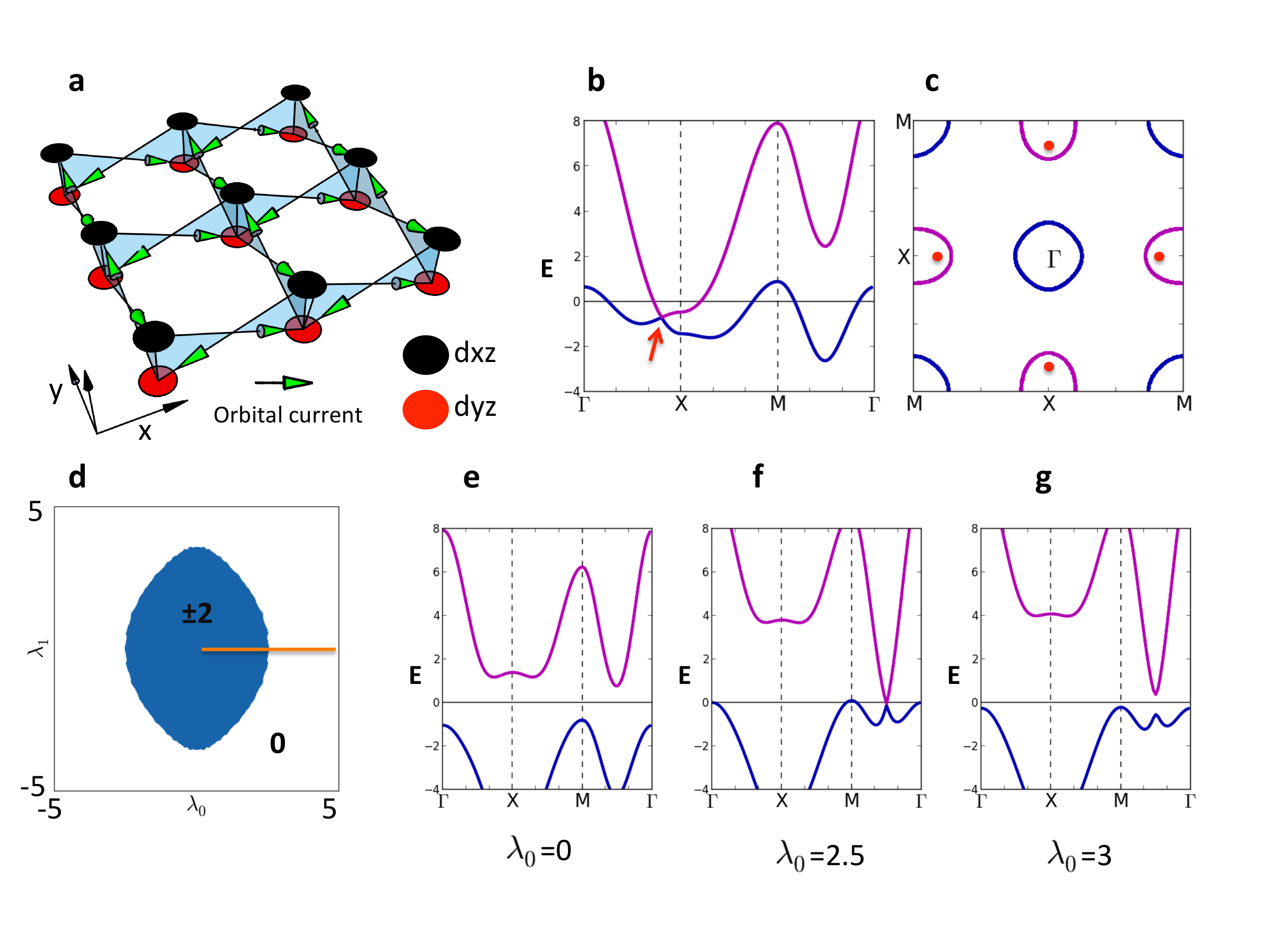}
\caption {{(color online) Inter-orbital currents, electronic structure and phase diagram.}
{(a)} The schematics of the orbital current order with inter-orbital Coulomb coupling $\lambda_{AOH}$.
The coordinates $x$ and $y$ are defined along the nearest bond directions.
{(b-c)} The band structure and Fermi surfaces in the 2D BZ at half filling.
The red-arrow(in b) and red-dot(in c) indicate the location of Dirac cone which underneath the Fermi surface.
{(d)} The calculated phase diagram is based on the Chern number
$\mathcal{C}_\epsilon^n$ with $\lambda_{AOH}=1$. 
{(e, f, g)}, The corresponding band structure
evolves from the Chern insulator to metal to trivial band insulator as function of $\lambda_0$ with $\lambda_{AOH}=1$ and  $\lambda_1=0$ along the orange line in the phase diagram of panel {(d)}.
}\label{pic:ModelBerry}
\end{figure}

\paragraph{The spinless $t_{2g}$ orbital model.$-$}
We start with a simplified $t_{2g}$ orbital model Hamiltonian, $H=H^0+H^{V}+h_{0}+h_{1}$, for spinless fermions to facilitate our symmetry analysis and discussion. Here $H^0$ is the hopping term, $H^{V}$ is the interaction term, and $h_{0,1}$ are perturbation terms:
\begin{equation}
\begin{aligned}
	H^0 =&\sum_{IJ,\alpha\beta} (t_{IJ}^{\alpha\beta}-\mu\,\delta_{IJ}\delta_{\alpha\beta})\;c^\dagger_{I\alpha}\,c_{J\beta} , \\
	H^{V}	=&i\,\epsilon\,\lambda_{AOH}\sum_{IJ,\alpha} \nu_{IJ}^{\alpha\bar\alpha}\; c^\dagger_{I\alpha} c_{J\bar\alpha} , \\
	h_0	=&\lambda_0\,\sum_{I,\alpha} (-1)^{\alpha} c^\dagger_{I\alpha}c_{I\alpha} , \\
	h_1	=&i\,\epsilon\,\lambda_1\,\sum_{I,\alpha} \, (-1)^\alpha\, c^\dagger_{I\alpha}c_{I\bar\alpha} ,
	\label{eq:Ham}
\end{aligned}
\end{equation}
where $I, J$ are lattice site indices, and $\alpha,\beta \in [1,2]$ are indices for the $d_{xz}$ and $d_{yz}$ orbitals in the $t_{2g}$ manifold.
We choose the nonvanishing hopping elements as $t^{\alpha\alpha}_{\pm\hat x} = t^{\alpha\alpha}_{\pm\hat y}=t_1$, $t^{11}_{\pm(\hat x+\hat y)} = t^{22}_{\pm(\hat x-\hat y)}=t_2$, $t^{11}_{\pm(\hat x-\hat y)} = t^{22}_{\pm(\hat x+\hat y)}=t_3$, $t^{\alpha\bar\alpha}_{\pm(\hat x\pm\hat y)}=t_4$, $t^{\alpha\bar\alpha}_{\pm\hat x} = t^{\alpha\bar\alpha}_{\pm\hat y}=t_5$, $t^{\alpha\alpha}_{\pm2\hat x}=t^{\alpha\alpha}_{\pm2\hat y}=t_6$ with $t_{1-6}=(0.09, 0.08, 1.35, -0.12, -1, 0.25)$.
The tensor elements $\nu_{IJ}^{\alpha\bar\alpha} \in [0,\pm 1]$ describe the direction of the NN inter-orbital currents as shown in Fig.~\ref{pic:ModelBerry}(a) with $\nu_{\pm\hat x}^{12}=\nu_{\pm\hat y}^{21}=-1$, and $\nu_{\pm\hat x}^{21}=\nu_{\pm\hat y}^{12}=1$.
The scalar $\epsilon=\pm 1$ describes the direction of the orbital current loop or the direction of the arrows in Fig.~\ref{pic:ModelBerry}(a).
The hopping parameters between different lattice sites and orbitals are given by $t^{\alpha\beta}_{IJ}$, and  $\mu$ is the chemical potential which includes the mean-field energy shift from the onsite Coulomb interaction.
The {\it anomalous orbital Hall} (AOH) effect is the multiorbital analog of the anomalous Hall effect and is described by the complex hopping term between different orbitals and different sites $I$ and $J$ with the spinless coupling constant $\lambda_{AOH}=V_{1}\, {\rm Im}|\langle c_{I\alpha}^{\dagger}c_{J\bar\alpha}\rangle|$ determined by the current order $\langle c_{I\alpha}^{\dagger}c_{J\bar\alpha}\rangle$ through the NN inter-orbital Coulomb interaction $V_{1}$.
The real part of the current order, $\delta t=-V_{1}\, {\rm Re}|\langle c_{I\alpha}^{\dagger}c_{J\bar\alpha}\rangle|$, can be absorbed into the hopping terms $t_{IJ}^{\alpha\beta}$ and does not affect our conclusions.
In this paper all the parameters are in units of $|t_5|$ which can be adjusted to fit the band structure from the ARPES experiment or the DFT calcuations.

The onsite orbital energy difference $\lambda_0$ in the term $h_0$ is responsible for the orbital charge polarization, which can be induced by an external electric field perpendicular to the lattice or anisotropic strain from the substrate.  
On the other side, the coupling constant $\lambda_{1}$ in $h_1$ is responsible for inter-orbital coherence.
Although $\lambda_0$ and $\lambda_1$ may be negligible in real systems, they allow us to perform a stability analysis of the topological phases toward TR symmetry violation.

Due to the translational invariance of the periodic lattice structure, the Hamiltonian $H$ can be diagonalized in the momentum space ${\bf k}$, that is, $H({\bf k})=\frac{1}{N}\sum_k \psi_{\bf k}^\dagger \hat H({\bf k}) \psi_{\bf k}$, with basis functions $\psi_{\bf k}=(c_{{\bf k},1},c_{{\bf k},2})^T$ (where $T$ is the transpose operation).
We derive the expression $\hat H({\bf k})=E_0({\bf k}){\rm \hat I}+\vec B({\bf k})\cdot\vec\tau$, where ${\rm \hat I}$ is the 2-by-2 unit matrix, $\vec B=(X,Y,Z)$ and $\vec\tau=(\tau_x,\tau_y,\tau_z)$ are the Pauli matrices. The ancillary functions $E_0$, $X$, $Y$ and $Z$ are given by
\begin{equation}
\begin{aligned}
	E_0 =& 2 t_1 [ \cos(k_x)+\cos(k_y) ] + 2 t_6 [ \cos(2k_x)+\cos(2k_y) ] , \\
		&+ 2 (t_2+t_3) [ \cos(k_x) \cos(k_y) ]-\mu , \\
	X	=& 4 t_4 [ \cos(k_x) \cos(k_y) ]+ 2 (t_5+\delta t) [ \cos(k_x)+\cos(k_y) ] , \\
	Y	=& \epsilon\,(\,2 \lambda_{AOH} [\cos(k_x)-\cos(k_y)]+ \lambda_1), \\
	Z	=& 2 (t_2-t_3) [ \sin(k_x) \sin(k_y) ]+ \lambda_0.
	\label{element}
\end{aligned}
\end{equation}
The diagonalization of $\hat H({\bf k})$ attains the eigenvalues  $E_\pm({\bf k}) = E_0({\bf k}) \pm B({\bf k})$, where $B=|\vec B|$. The corresponding eigenvectors are
\begin{equation}
\begin{aligned}
	|+,{\bf k}\rangle &= (Z+B, X+i\, Y)^T/\sqrt{2B^2+2ZB},\\
	|-,{\bf k}\rangle &= (-X+i\,Y, Z+B)^T/\sqrt{2B^2+2ZB}.
\end{aligned}
\end{equation}
We find an even number of (four) Dirac cones in the dispersion of the noninteracting bulk material, i.e., $\lambda_{AOH}=\lambda_0=\lambda_1=0$.
Their positions are located at the $k_{x}$ and $k_{y}$ axes as determined by $B({\bf k})=0$, see Fig.~\ref{pic:ModelBerry}(b) ~\cite{footnote3}.
For any finite orbital current order ($\lambda_{AOH}>0$) a nonzero $Y$ will be generated, inducing the anomalous orbital Hall effect.
Consequently, the Dirac cones in the bulk, which are responsible for nontrivial band topology, become gapped. If we manually turn off $Z$ in the presence of $Y\neq 0$, the four Dirac cones are pushed toward the center of each quadrant of the BZ. When spin degrees of freedom are considered, the term $Y$ is also responsible for band topology protected by reflection symmetry, $\sigma_v$, which is the main focus of this work.

It is worth to note that similar ideas about the importance of the NN-inter-orbital Coulomb interaction and resulting current flux phases have been discussed in the context of bilayer graphene \cite{LZhu2013, XZYan2012, XZYan2014} and the cuprates \cite{Varma1999,Sudip,JXZhu,Simon}.
Although the origin of our proposed inter-orbital currents 
is similar to the loop or circulating current flux phase in the pseudo-gap phase of the cuprates~\cite{Varma1999,Sudip,JXZhu,Simon}, 
our model does not rely on strong onsite Coulomb interaction and has different orbital degrees of freedom, crystal symmetry, and conduction band topology. 
Specifically, the single-orbital models of the cuprates with $d$-density-wave order~\cite{Sudip,JXZhu} break the 1-atom per unit cell translational invariance and the TR symmetry. The current loop model~\cite{Simon} violates the TR symmetry.

\paragraph{Anomalous Orbital Hall Phases.$-$}
The ground state of the spinless Hamiltonian $H$ is illustrated in Fig.~\ref{pic:ModelBerry}(a) based on standard self-consistent mean-field calculations.
In Fig.~\ref{pic:ModelBerry}(b) and \ref{pic:ModelBerry}(c) we show the dispersion of the electronic band structure of the noninteracting bulk material at half filling when $\lambda_{AOH}=\lambda_0=\lambda_1=0$.
For insulators, the nonlocal topology of band $n$ can be captured by the Chern number 
$\mathcal{C}^n_\epsilon$ directly through the Berry curvature $\Omega^n({\bf k})$ 
of the Hamiltonian $\hat H({\bf k})$, which is defined as $\mathcal{C}^n_\epsilon\,(\epsilon;\lambda_{AOH};\lambda_0;\lambda_1) =\frac{1}{2\pi}$$\int_{{\bf k}\in BZ} dk_{x}dk_{y}\, \Omega^n({\bf k})$, where the expression of $\Omega^n({\bf k})$ is defined in the Supplemental Material (SM) \cite{SM}.
The first observation is that $\mathcal{C}^n_\epsilon\,(\lambda_{AOH};0;0)=\pm2$ for any real but nonzero $\lambda_{AOH}$.
The second observation is that the sign of $\mathcal{C}^n_\epsilon$ depends on the direction of the orbital current loop through $\epsilon$ and the band index $n\in[1,2]$, which determine the class of the Chern insulator, $\mbox{sign}(\,\mathcal{C}^n_\epsilon\,)=\epsilon\times(-1)^n$.
Since the topological phase with $\mathcal{C}^n_\epsilon = \pm2$ is robust against weak perturbations by TR symmetry violation, we show its stability region in the $\lambda_0$-$\lambda_1$ phase diagram in Fig.~\ref{pic:ModelBerry}(d), where  for illustration purposes we chose the strong coupling limit $\lambda_{AOH}=1$. 
Note that the TI phase is induced by interaction and therefore vanishes for $\lambda_{AOH} \to 0$.
Following the orange line in the phase diagram, we monitor the evolution of the bulk band gap as it closes and reopens with increasing  $\lambda_0$, see Figs.~\ref{pic:ModelBerry}(e) to \ref{pic:ModelBerry}(g).
This leads to a sequence of phase transitions from topological
Chern insulator
to metal (around ${\lambda_{0}\approx 2.5}$) and on to trivial band insulator with $\mathcal{C}^n_\epsilon = 0$.

\begin{figure}
\includegraphics[scale=0.35,angle=0]{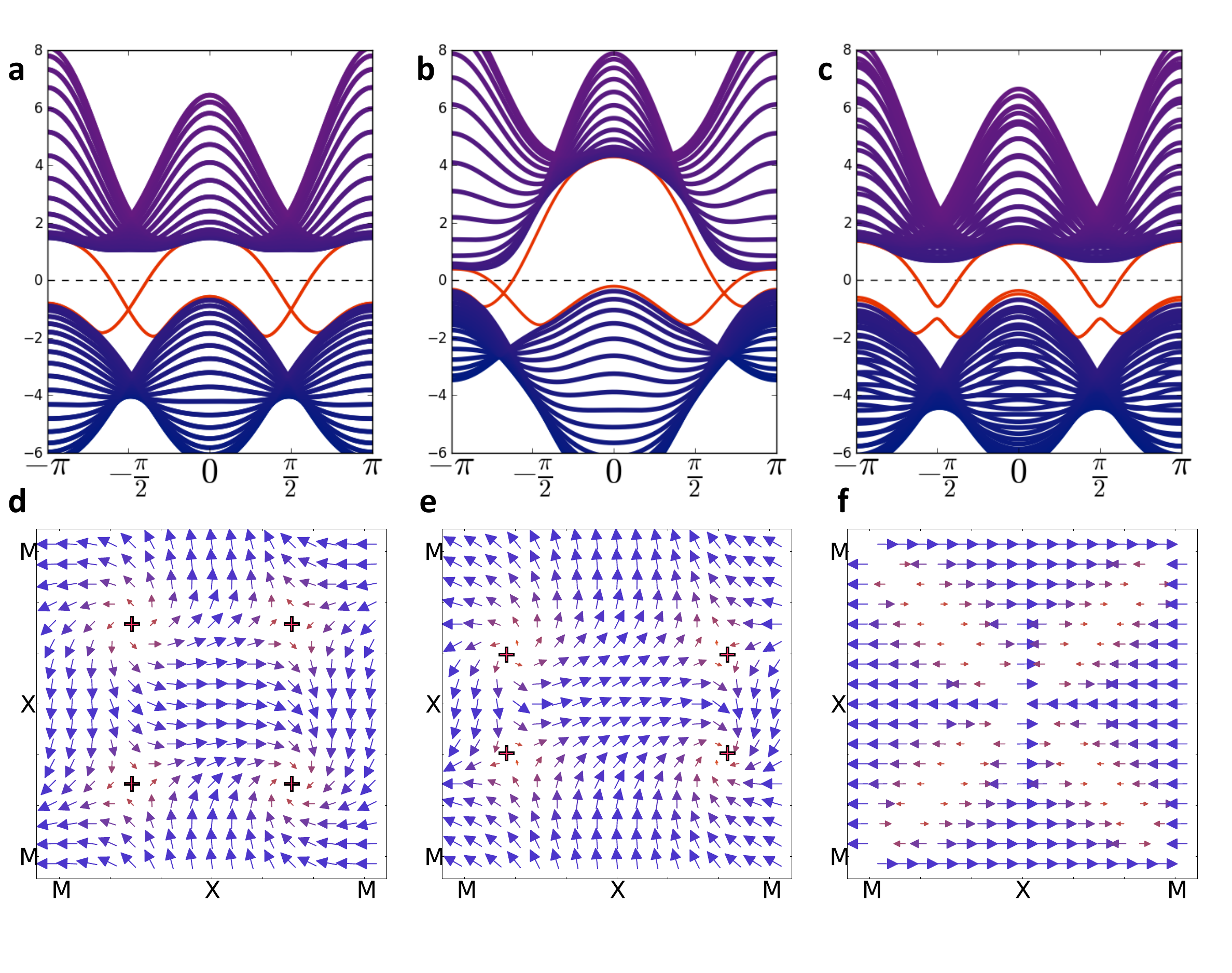}
\caption
{{(color online) Subbands, edge states with two-fold spin degeneracies in a strip geometry and vector plots of Pfaffian in two-dimensional BZ.}
	{(a-c)} The subbands including four edge states (red lines) for a strip with 20 lattice sites in open boundary width and 100 k-points along the periodic boundary direction.
Different parameters with fixed Coulomb coupling $\lambda_{AOH}=1$ are shown as illustrated as follows:
	{(a)} TMI with $\lambda_0=\lambda_1=\lambda_R=0$;
	{(b)} TMI with $\lambda_1=2$, $\lambda_{0}=\lambda_{R}=0$;
	{(c)}  Band insulator with $\lambda_{R}=0.03$, $\lambda_{0}=\lambda_{1}=0$. Note that any finite Rashba coupling $\lambda_{R}$ splits the Dirac cones.
	{(d-f)} The corresponding vector plots of the complex Pfaffian function $\mathcal{P}({\bf k})$ in two-dimensional periodic boundary conditions.
	The bi-color code represents small (orange) to large (blue) modulus of $|\mathcal{P}({\bf k})|$.	
	The red crosses mark the positions of the vortex cores where $\mathcal{P}({\bf k})=0$.
}\label{pic:EdgeVortex}
\end{figure}

\paragraph{The spinfull $t_{2g}$ orbital model.$-$}
In materials with magnetic interactions we need to consider electrons as fermions with spin degrees of freedom.  Therefore, we promote the spinless two-band orbital model to the spinful model ${H}_{s}={H}_\uparrow[\epsilon_\uparrow]+{H}_\downarrow[\epsilon_\downarrow]$.
Here the sign of the spinful orbital current direction is denoted as $\epsilon_{\sigma}=\pm 1$ for each spin index $\sigma\in [\uparrow, \downarrow]$.
A detailed analysis of the Hamiltonian ${\hat H}^s$ (see the SM) reveals the following invariants of the stable topological phases: \\
\;\;$\bullet$\;Phase~{\bf I}: \;\;$\epsilon_\uparrow = \epsilon_\downarrow$ with Chern number $\mathcal{C}=\pm4 \ (\mathcal{C}^n_{\uparrow}=\mathcal{C}^n_{\downarrow})$,\\
\;\;$\bullet$\;Phase~{\bf II}: \,$\epsilon_\uparrow=-\epsilon_\downarrow$ with Chern number $\mathcal{C}=0 \, (\mathcal{C}^n_{\uparrow}=-\mathcal{C}^n_{\downarrow})$.\\
For phase {\bf  I}, we find that the Chern number $\mathcal{C}=\pm4$ of the occupied bands is twice that of the spinless case due to the twofold degeneracy of spins, because degenerate spins share the same orbital current direction.
For phase {\bf  II}, we find that the Chern number classification scheme is insufficient to capture the topological nontrivial insulator phase, because of $\mathcal{C}=0$.

It is interesting to note that the form of the interaction term in phase {\bf II} is formally equivalent to an inter-orbital SOC, 
$H^V_s=i\lambda_{AOH}\sum_{IJ\alpha\beta} c^\dagger_{I\alpha}\,(\nu^{\alpha\beta}_{IJ}\hat z\cdot  \vec\sigma)\, c_{J\beta}$ with $c_{I\alpha}=(c_{I\alpha\uparrow},c_{I\alpha\downarrow})^T$.
This equivalence shows that a distinction between correlation- and SOC-induced topological states may not be that important after all, and similar analogies for the interaction driven phase to the SOC has already been discussed in other topological systems~\cite{KYYang,ARuegg}.
Here, if we regard this term as an intrnsic SOC and interplay with the NN inter-orbital Coulomb interaction ($V_1$), this results the imaginary part of the orbital order emerges earlier as $V_1$ increase.

To see whether phase {\bf II} is protected by band topology, we plot the edge states of the slab geometry
in Fig.~\ref{pic:EdgeVortex}(a).
The calculated edge states along the (1,0) direction show two surface Dirac cones at $k_x$=$\pm\frac{\pi}{2}$.
Furthermore, these edge states are robust against the TR perturbation $\lambda_1$ up to a critical value of roughly 3, although the position of the surface Dirac cones evolves away from $k_x=\pm\frac{\pi}{2}$ (see also Fig.~\ref{pic:EdgeVortex}(b) for $\lambda_1=2$).
One may tend to claim that phase \textbf{II} of the TMI is a conventional $Z_{2}$ quantum-spin Hall insulator, since the TR symmetry is respected by the mean-field Hamiltonian for phase \textbf{II} with $\mathcal{C}=0$. However, this cannot be reconciled with the fact that in our case the number of pairs of degenerate edge states is even instead of odd, as is the case for the quantum spin-Hall insulator.
Consequently, we claim that phase \textbf{II} has topology different from earlier work~\cite{Kane} and is a new type of topological phase in 2D, protected by mirror reflection symmetries (spinful). This is accomplished by TR symmetry (spinful) and reflection symmetry (spinless), as indicated by the even mirror Chern number $\mathcal{C}_{M}$~\cite{Fu3,Teo}.  $\mathcal{C}_{M}$ is related to the spin Chern number of the occupied band with spin up/down,
$\mathcal{C}^1_{\uparrow,\downarrow}$, and given by $|\mathcal{C}_{M}|=|(\mathcal{C}^1_{\uparrow}-\mathcal{C}^1_{\downarrow})/2| =[2-(-2)]/2=2$, 
as opposed to the $Z_{2}$ quantum-spin Hall insulator in the Kane-Mele lattice model with odd  mirror Chern number $\mathcal{C}_{M} = 1$.
In the next section, we propose a mirror Pfaffian with a $Z_2$ invariant to connect the nontrivial topology of the TI protected states with their mirror symmetries in phase \textbf{II}.

\paragraph{Mirror-Z$_2$ topological invariant.$-$}
The spinful mean-field Hamiltonian of phase \textbf{II} respects the TR symmetry. However, the number of degenerate Dirac cone pairs at the edges is even instead of odd in addition to an even number of Dirac cones in the bulk BZ. This is in sharp contrast to quantum-spin Hall insulators, which are solely protected by the TR symmetry and other spatial symmetries such as inversion symmetry.
A detailed symmetry analysis (see the SM) reveals that the topological phase \textbf{II} of the TMI satisfies the mirror symmetry under the combination of {\it space} ($\sigma_v$) and {\it time} (spin) operations, 
$\mM{\hat H}_{s}(k_x,k_y)\mM^{-1}={\hat H}_{s}(\pm k_x,\mp k_y)$, 
in which the mirror operator is given by $\mM=\mP\otimes\mT= (\tau_{x}\mK)\otimes(-i\sigma_{y}{\mK})=\tau_x\otimes i\sigma_{y}$.
The operator $\mK$ performs the complex conjugation identical to the TR operation for spinless fermions.
The generalized parity operator $\mP$ exchanges two orbitals, while $-i\sigma_{y}$ is responsible for the spin flip under the TR operation $\mT$.
The overall $\mM$ operation is equivalent to mirror reflection (including spin sectors) with respect to the principle axis $k_x=0$ or $k_y=0$.

Analogous to the analysis in the Kane-Mele model for the quantum-spin Hall insulator~\cite{Kane}, we introduce a mirror-invariant Pfaffian for occupied states
to quantify the $Z_{2}$ invariant of the ``even/odd parity" of the spinful Hamiltonian ${\hat H}_{s}({\bf k})$ with the mirror symmetry $\mM$.
Specifically, we  define the mirror-invariant Pfaffian to measure the band topology as
\begin{equation}
\begin{aligned}
	\mathcal{P}({\bf k})& \equiv Pf\big[ \langle u_m({\bf k})| \mM |
u_n({\bf k}) \rangle\big],
\end{aligned}
\end{equation}
where $|u_m({\bf k})\rangle,|u_n({\bf k})\rangle$ are two occupied orthogonal eigenstates of the Hamiltonian $\hat H_s({\bf k})$, e.g., $n=1$ and $m=2$ or vice versa.
The commutation relation $[ \mM, \hat H_{s}({\bf k}) ]=0$ holds
for $\bf k$ points belonging to the  ``even parity" subspace along the boundaries of the four quadrants of the BZ.
Therefore, the two occupied eigenstates $\mM|u_{n}({\bf k})\rangle$ and $|u_{n}({\bf k})\rangle$ are identical states up to a phase factor.
As a result, the absolute value of the Pfaffian $\mathcal{P}({\bf k})$, with ${\bf k}$ along the $k_{x}$  and $k_{y}$ axes is unity, $|\mathcal{P}({\bf k})|=1$.
On the other hand, $\bf k$ points belonging to the ``odd parity'' subspace, 
given by the roots of the Pfaffian,
satisfy the anti-commutation relation $\{ \mM , \hat H_{s}({\bf k}) \} = 0$.
Here the mirror operation $\mM|u_{n}(\bf k)\rangle$ turns one occupied state, for example, at ${\bf k} =(\pi/2,\pi/2)$ into an unoccupied and orthogonal eigenstate at ${\bf k} =(-\pi/2,\pi/2)$, $|u_{m}({\bf k})\rangle$ and vice versa, with vanishing Pfaffian $\mathcal{P}({\bf k})=0$ for the occupied states at ${\bf k} =(\pi/2,\pi/2)$.

In Fig.~\ref{pic:EdgeVortex}(d) we show four vortices appearing in the Pfaffian for phase \textbf{I}I with opposite vorticity in adjacent quadrants of the BZ. All four vortices are well separated by the ``even parity'' subspace along the $k_{x}$ and $k_{y}$ axes or the boundaries of the BZ quadrants.
Note that for the TMI the even parity subspaces are connected lines which is different from the case of  the TI with inversion symmetry~\cite{Fu}, where the even parity subspaces are separted points in the BZ.

It is an important question to confirm whether the proposed topological phase is protected by mirror symmetry. For that purpose, we examine the effects of a mirror-symmetry breaking perturbation on $\mathcal{P}({\bf k})$.
To perform a stability analysis, we introduce an onsite SOC interaction, which might be called an onsite Rashba term, 
$h_R=i\lambda_R\,\sum_{I\alpha\sigma} (-1)^{\alpha}(-1)^\sigma c^\dagger_{I\alpha\sigma} c_{I\alpha\bar\sigma}$,
but is different in nature from the usual off-site Rashba term:
The corresponding matrix elements, written in matrix notation in momentum space as $\hat h_R({\bf k})=\lambda_R\,\tau_z\otimes\sigma_y$, break both the TR and space-time mirror symmetry. In other words, together with $\hat H_s({\bf k})$, $h_R$ does not commute with $\mM$ anywhere in the BZ, $[ \mM, \hat H_s({\bf k}) + \hat h_R({\bf k}) ]\neq 0$.
This symmetry breaking field will destroy the mirror topological phase even though the interaction $\hat h_R({\bf k})$ is infinitesimal.
As we expect, the four vortices (Dirac cones) disappear for any nonzero onsite Rashba-like SOC interaction as shown in Fig.~\ref{pic:EdgeVortex}(f). Consequently, an infinitesimal $\lambda_R$ destroys the degeneracy of edge states and the previously gapless (crossing) edge states become gapped, see Fig.~\ref{pic:EdgeVortex}(c).

A completely different scenario occurs when the local inter-orbital coupling $\lambda_{1}\tau_{y}$ is  turned on adiabatically.
For this case, the Pfaffian is plotted in Fig.~\ref{pic:EdgeVortex}(e).
As the strength of $\lambda_1$ increases the positions of the pair of vortices in the upper half-plane of the BZ are modified and move toward the pair in the lower half-plane compared to the onsite SOC case in Fig.~\ref{pic:EdgeVortex}(d).
As we already mentioned before, this trend continues until the vortices disappear at a critical strength $\lambda_{1}\approx 3$ before entering the even parity subspace protected by the mirror symmetry. Indeed this corresponds to the stability boundary discussed previouly in the phase diagram in Fig.~\ref{pic:ModelBerry}(d) of spinless fermions.
Furthermore this observation is consistent with the corresponding degeneracies of edge states as displayed in Fig.~\ref{pic:EdgeVortex}(b).
Therefore, according to the mirror symmetry, which maps the entire {\bf k} space of the BZ onto one quadrant, a new mirror-Z$_2$ index can be defined to count the number of vorticies of the Pfaffian in one quadrant of the BZ (see the SM).

\paragraph{Conclusion.$-$}
Our work shows that unconventional topological insulators can emerge from Coulomb correlations in real materials with non-local crystal symmetry in the absence of spin-orbit coupling.
A remarkable result of our quasi-2D model Hamiltonian is the finding of an even number of pairs of Dirac cones at the edges. 
In the spinless case, the four edge states are described by the topological Chern number $\mathcal{C}^n_\epsilon = \pm 2$.
On the other hand, in the spinful case of phase \textbf{I}  the Chern number is $\mathcal{C} = \pm 4$,
while in phase \textbf{II}  the Chern number $\mathcal{C}=0$ is insufficient to classify the topology. 
In that case, the Pfaffian enumerates the four edge states and is connected to a $Z_2$ invariant.
Similar to the previous Z$_2$ invariant in TIs with inversion symmetry~\cite{Kane,Fu}, the mirror-Z$_2$ invariant in TMI is robust against TR breaking perturbations, because the topological state is protected by a mirror reflection symmetry.

Similar ideas of the mirror-Chern number \cite{Fu3,Teo} and the mirror reflection symmetry of the $C_{nv}$ group \cite{AA} have been discussed before.
The novalty here is that we have extended these cases to a spinfull Hamiltonian in the absence of spin-orbit coupling and  found a new mirror-Z$_2$ index in phase {\bf II}.

Finally, our studies provide a new direction toward the realization of correlation-induced topological phases in $d$-orbital material.
We suggest to search for TMIs in the paramagnetic iron-pnictide compounds with crystallographic 11, 111, 122 and 1111 structures near half-filling \cite{Paglione}, where the $t_{2g}$ model is expected to be valid.
In view of recent interest in superconducting topological phase~\cite{DHLee}, the interplay of our orbital order proposed here and superconductivity will be of a very interesting topic for future study.

\acknowledgments
\paragraph{Acknowledgements.$-$}
We acknowledge Fan Zhang, J. Ren, G.-W. Chern and Tanmoy Das for sharing their views on topological insulators.
The work at Houston was supported in part by the Robert A.\ Welch Foundation under Grant No.\ E-1146 and AFOSR under Grant No. FA9550-09-1-0656.
The work at Los Alamos was supported by the U.S.\ DOE Contract No.~DE-AC52-06NA25396 through the
LDRD program (Y.-Y.T., C.-C.J.W.), the Office of Basic Energy Sciences (BES), Division of Materials Sciences and Engineering (M.J.G., J.-X.Z.). This work was supported in part by the Center for Integrated Nanotechnologies, a DOE BES user facility.

\section{Supplemental Material: Emergent topological mirror insulator in $t_{2g}$-orbital systems}
%\pacs{78.70.Dm, 71.10.Fd, 71.10.-w, 71.15.Qe}

%-----------------------------------------------------------------------------------------
\maketitle

In this Supplemental Material, we provide the additional technical information and details used in the main text of the publication.
The following sections contain the in-depth description of the model, supporting numerical calculations, and symmetry analysis:

\begin{enumerate}
\item Crystal field splitting of $d$ orbitals
\item Mean-field lattice Hamiltonian of correlated electron system
\item Inter-orbital current order of the flux phase
\item Lifting of ground state degeneracy with exchange interaction
\item The Hamiltonian in momentum representation
\item Two types of orbital order with C$_{4v}$ symmetry
\item Vortices as generators of the Berry flux and Chern numbers
\item Symmetry analysis of the time and mirror invariance
\end{enumerate}

\section{Crystal field splitting of $d$ orbitals}
\begin{figure}
\includegraphics[scale=0.3,angle=0]{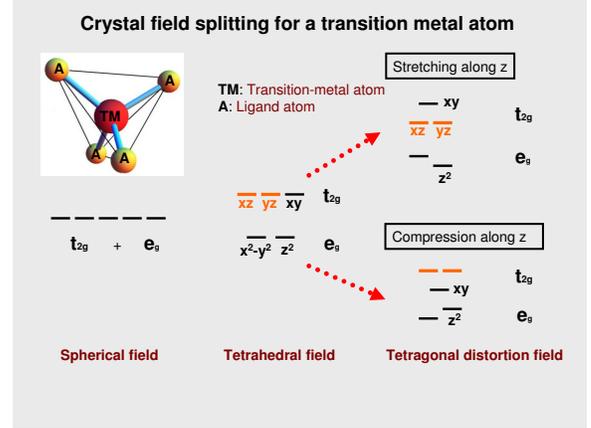}
\caption {(color online) {\bf Crystal field splitting.}
In a spherical crystal field, the five $d$ orbitals of the TM atom including $d_{xy}, d_{xz}, d_{yz}, d_{x^2-y^2}$, and $d_{z^2}$ are degenerate.
With a tetrahedral crystal field, the five $d$ orbitals into $t_{2g}$ and $e_{g}$ manifolds where $t_{2g}$ levels are higher in energy. With a tetragonal crystal field distortion, part of the degeneracy of the $t_{2g}$ orbitals is lifted with the $d_{xy}$ level energetically separated from $d_{xz}$ and $d_{yz}$ orbitals. When the strain is stretched (compressed) along the layer growth direction, the wave function overlaps between $d_{xz}$ and $d_{yz}$ orbitals are reduced (enhanced)
causing the reduction (enhancement) in Coulomb interaction. Therefore, $d_{xz}$ and $d_{yz}$ are lower (higher) in energies.
}\label{pic:CF}
\end{figure}

For a transition-metal (TM) atom situated in crystal fields due to
surrounding ions in compounds, the $d$ atomic levels align differently. The five $d$ orbitals include $d_{xy}, d_{xz}, d_{yz}, d_{x^2-y^2}$, and $d_{z^2}$. In a spherical crystal field, the $d$ orbitals are all degenerate. For a TM atom caged by four ligand atoms, a tetrahedral crystal field splits the five $d$ orbitals into $t_{2g}$ and $e_{g}$ manifolds, where $t_{2g}$ levels are higher in energy. With a tetragonal crystal field distortion, introduced by anisotropic strains or effective strains with different types of the surrounding ligand atoms, part of the degeneracy of the $t_{2g}$ orbitals is lifted with the $d_{xy}$ level energetically separated from $d_{xz}$ and $d_{yz}$ orbitals, which are the relevant degrees of freedom we are interested in. \cite{Khomskii}
The $d_{xy}$ level will be higher in energy for uniaxial stretching strain along the $z$ axis and lower for compression as shown in Fig.~(\ref{pic:CF}).

\section{Mean-field lattice Hamiltonian of correlated electron system} 
In this section, we derive the mean-field form of the Coulomb interaction.
The Hamiltonian of interest is $H^s=H^0 + H^U +H^J + H^V$, where the superscript  $s$ indicates that we include the spin degrees of freedom.
In real-space the lattice Hamiltonian is given by
\begin{equation}
\begin{aligned}
	H^0	&=\sum_{IJ,\alpha\beta,\sigma} (t_{IJ}^{\alpha\beta}-\mu\,\delta_{IJ}\delta_{\alpha\beta}) c^\dagger_{I\alpha,\sigma} c_{J\beta,\sigma},\\
	H^U &=U\sum_{I,\alpha,\sigma} \; n_{I\alpha,\sigma}\, n_{I\alpha,\bar\sigma},\\
	H^J_h	 &=U'\sum_{I\sigma\alpha}\;n_{I\alpha,\sigma}n_{I\bar\alpha,\bar\sigma} +(U'-J_h)\sum_{I\sigma\alpha}\;n_{I\alpha,\sigma}n_{I\bar\alpha,\sigma},\\
	H^V &=\sum_{I\neq J,\alpha,\sigma} V_{IJ}\; n_{I\alpha,\sigma}\, n_{J\bar\alpha,\sigma},
	\label{Ham}
\end{aligned}
\end{equation}
where $U'=U-2J_h$.
Here $H^0$ is the kinetic term describing the hopping of electrons in the 1-Fe per unit cell formulation.\cite{YTai,HChen}
The ``bar" above subscripts stands for the opposite orbital or spin component, i.e.,  $\bar\alpha \neq \alpha$ and $\bar\sigma \neq \sigma$.
In our two-orbital model with $\alpha, \beta =(1,2)$  the non-zero hopping parameters are chosen to describe the
generic pnictide BaFe$_2$As$_2$ for purpose of illustration,
\begin{equation}
\begin{aligned}
t_1=\,&t^{\alpha\alpha}_{\pm \hat{x}}=t^{\alpha\alpha}_{\pm \hat{y}}=0.09,\\
t_2=\,&t^{11}_{\pm (\hat{x} - \hat{y})}=t^{22}_{\pm (\hat{x}+\hat{y})}=0.08,\\
t_3=\,&t^{11}_{\pm (\hat{x} + \hat{y})}=t^{22}_{\pm (\hat{x}-\hat{y})}=1.35,\\
t_4=\,&t^{\alpha\bar{\alpha}}_{\pm (\hat{x} \pm \hat{y})}= -0.12,\\
t_5=\,&t^{\alpha\bar{\alpha}}_{\pm \hat{x}}=t^{\alpha\bar{\alpha}}_{\pm \hat{y}}=-1,\\
t_6=\,&t^{\alpha\alpha}_{\pm 2\hat{x}} = t^{\alpha\alpha}_{\pm 2\hat{y}}=0.25.\\
\end{aligned}
\end{equation}

\begin{figure}
\includegraphics[scale=0.55,angle=0]{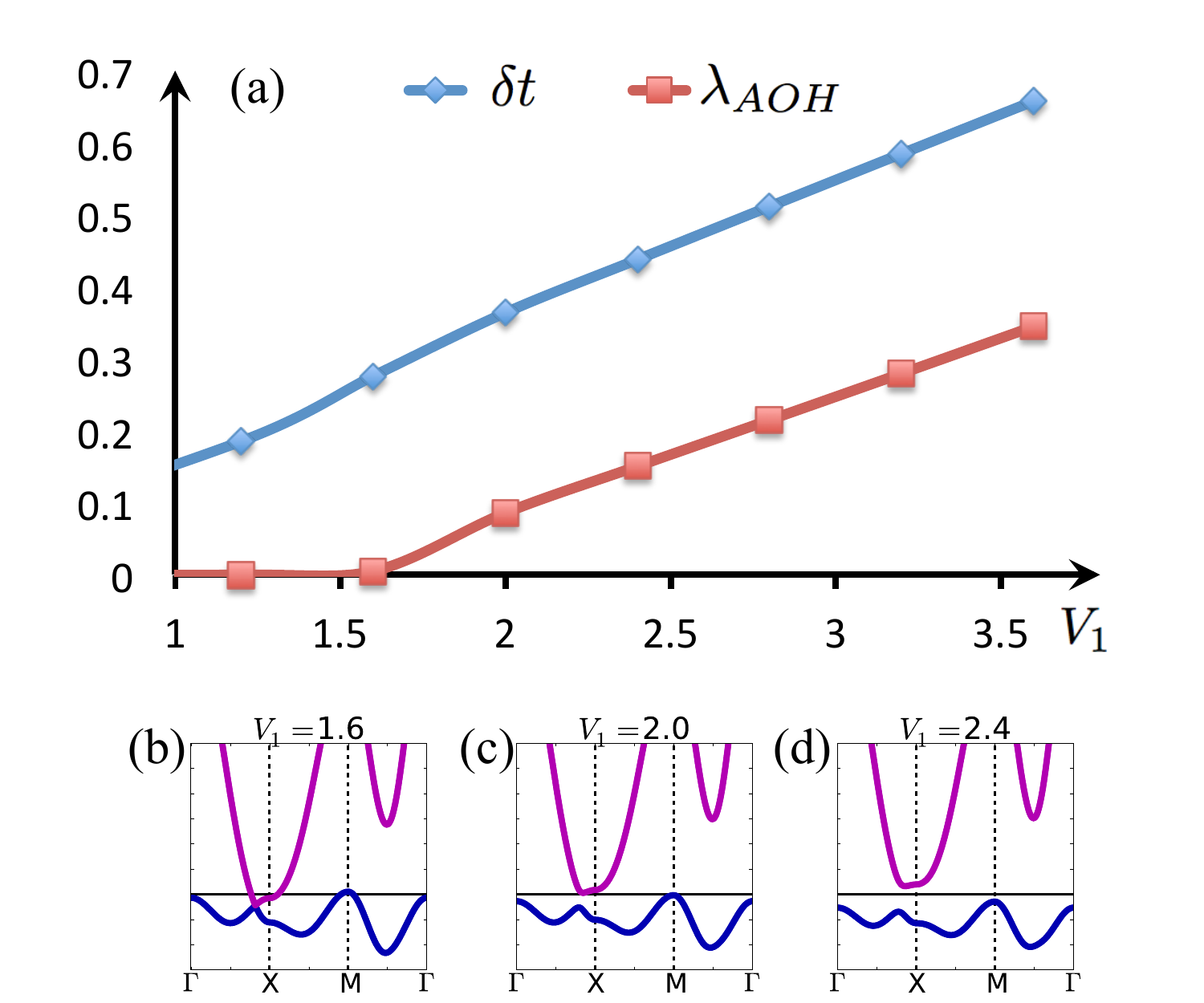}
\caption { (color online) (a) The mean-field order parameter of the $H^{AOH}$ term, where $\delta t$/$\lambda_{AOH}$ are the real/imaginary part of the order parameter. (b),(c) and (d) The evolution of the dispersion of the electronic band structure for the 1-Fe/unit cell along high-symmetry directions in the BZ for different interaction strengths of $V_{1}$.
}\label{pic:Flux}
\end{figure}

The interaction part is captured by the terms $H^U$ and $H^J$, which are the on-site intra-orbital {\it Hubbard interaction} and {\it Hund's coupling},  as well as the term  $H^V$, which is the inter-orbital ($\alpha\neq\beta$) {\it Coulomb} interaction between lattice sites $I$ and $J$.
Note that we also considered  intra-orbital offsite Coulomb interaction ($\alpha=\beta$), but found no interesting topological phases.
Thus, we will not further consider the intra-orbital interaction in the current work.
By investigating $H^U$ and $H^V$ in the mean-field approximation, we can test whether there exist new and anomalous ground states due to the inter-orbital Coulomb interaction, although the incorporation of quantum fluctuations may change the details of such a phase diagram.
For the on-site inter-orbital $H^U$ we write in standard mean-field approximation
\begin{equation}
	H^U = U\, \sum_{I\alpha,\sigma\neq\sigma'} \langle n_{I\alpha\sigma} \rangle\,n_{I\alpha\sigma'}.
\end{equation}
On the other hand, we have at least two possibilities for the inter-orbital $H^V$ to decouple the fermionic operators within mean-field theory, namely $H^V = H^{CDW}+H^{AOH}$, where
\begin{equation}
\begin{aligned}
	H^{CDW}	&=\;\;\sum_{I\neq J,\alpha,\sigma} V_{IJ}\; \langle n_{I\alpha\sigma} \rangle\,n_{J\bar\alpha\sigma} ,\\
	H^{AOH} &=-\sum_{I\neq J,\alpha,\sigma} V_{IJ} \langle c^\dagger_{I\alpha,\sigma}\, c_{J\bar\alpha,\sigma} \rangle c^\dagger_{J\bar\alpha,\sigma}\, c_{I\alpha,\sigma} .\\
\end{aligned}
\end{equation}
In this work, we simplify the Coulomb coupling and include only the nearest-neighbor  (NN) interaction with $V_{\langle ij\rangle}$=$V_{1}$.
In the 2D-periodic calculation of the bulk material, the CDW term is not a stable ground state and only $H^{AOH}$ has a stable solution at a finite value of $V_{1} \agt 1.6$, see Fig.~\ref{pic:Flux}.
Unlike the Bardeen-Cooper-Schrieffer (BCS) theory, where any non-zero pairing strength will lead to superconductivity, here a threshold has to be overcome to induce long-range orbital order. Naturally, this makes it more challenging to find materials with orbital-ordered ground states.
We also checked numerically for magnetism and found that a Hubbard term with $U=3.2$ and $V_1\gg J$ does not induce long-range magnetic order in our simple, low-energy two-orbital model. 
While one might expect for real materials that the inter-site Coulomb interaction is less than the onsite interaction, $V_1 <  U$, we consider for illustrational purposes of the anomalous orbital effect the opposite case, when we show results for strong coupling with $\lambda_{AOH}=1$.
This also corresponds to the region of the interaction $V_1-J$ phase diagram where no magnetic order emerges.

%%%%%%%%%%%%%%%%%%%%%%%%%%
%%%%%%%%%%%%%%%%%%%%%%%%%

\section{Inter-orbital current order of the flux phase}
Here, let us define  the mean-field  inter-orbital current  order parameter of the flux phase due to the anomalous orbital Hall term $H^{AOH}$,
\begin{equation}
	\chi_{IJ}^{\alpha,\sigma}\equiv V_{1}\langle c^\dagger_{I\alpha,\sigma} c_{J\bar\alpha,\sigma} \rangle .
\end{equation}
The mean-field calculated nearest-neighbor order parameter $\chi_{IJ}^{\alpha\bar\alpha,\sigma}$ is a complex number with real and imaginary parts, $\chi_{IJ}^{\alpha\bar\alpha,\sigma}=\delta t+i\epsilon_\sigma\epsilon_{\alpha\bar\alpha}^{IJ}\,\lambda_{AOH}$, where we define,
\begin{equation}
\begin{aligned}
	\delta t &=-|\mbox{Re}\, \chi_{IJ}^{\alpha\bar\alpha,\sigma}|
	,\\
	\lambda_{AOH} &= |\mbox{Im}\, \chi_{IJ}^{\alpha\bar\alpha,\sigma}|.
	\label{chi}
\end{aligned}
\end{equation}
Note that the first term, $\delta t$, is always negative and homogeneous in real space.
It will contribute to $H^0$ through its NN inter-orbital hopping term $t_5$.
The second term, $\lambda_{AOH}$, is the generator of the anomalous orbital Hall effect, which is of key interest in this work.
Consequently, the term $H^{AOH}$ can be written in a more compact form,
\begin{equation}
	 H^{AOH}=i\,\epsilon_\sigma\,\lambda_{AOH}\,\sum_{IJ,\alpha,\sigma}
	 \nu_{IJ}^{\alpha\bar\alpha}\,\; c^\dagger_{I\alpha,\sigma} c_{J\bar\alpha,\sigma},
\end{equation}
where $\epsilon_\sigma = \epsilon_{\uparrow/\downarrow} = \pm1$ and the elements of the tensor  $\nu_{IJ}^{\alpha\bar\alpha} \in [0,\pm 1]$ describe the direction of the NN inter-orbital currents as shown in Fig.~1(a) of the main text with $\nu_{\pm\hat x}^{12}=\nu_{\pm\hat y}^{21}=-1$, and $\nu_{\pm\hat x}^{21}=\nu_{\pm\hat y}^{12}=1$.
The real part of $H^{AOH}$ has been absorbed into the NN inter-orbital hopping terms, $t_5 \rightarrow t_5+\delta t$.
Figure~\ref{pic:Flux}(b) shows that the hole pockets are shifted downward at the $\Gamma$ and $M$ points, due to the contribution of $\delta t$ alone.
Once the purely imaginary part $\lambda_{AOH}$ is included, the degeneracy of the bulk Dirac cone near the $X$ point in the Brillouin zone (BZ) is lifted immediately, as shown by the opening of a gap in Fig.~\ref{pic:Flux}(c).
The gap increases with increasing value of $\lambda_{AOH}$ as further shown in Fig.~\ref{pic:Flux}(d).

\section{Lifting of ground state degeneracy with exchange interaction}
In the main text, we have shown that without introducing other terms,  the ground state of $H^{AOH}$ leads to two degenerated phases, namely
$\epsilon_\uparrow=\pm\epsilon_\downarrow$.
In order to discuss the possibility of lifting this degeneracy, we introduce a perturbation caused by an additional Hund's coupling term,
to Eq.~(\ref{Ham}), as discussed by Sano\cite{Sano}, $H^J \rightarrow H^J+H^{J_h}_2$.
\begin{equation}
	H^{J_h}_2 = -J_h\sum_{I\alpha} (c^\dagger_{I,\alpha,\uparrow} c_{I,\alpha,\downarrow}\, c^\dagger_{I,\bar\alpha,\downarrow} c_{I,\bar\alpha,\uparrow}+h.c) .
\end{equation}
Then we have the mean-field decoupled Hamiltonian $H^{J_h}_2$,
\begin{equation}
	H^{J_h}_2 = \lambda_{J_h}\,\sum_{I\alpha} (c^\dagger_{I,\alpha,\uparrow} c_{I,\alpha,\downarrow}+ c^\dagger_{I,\bar\alpha,\downarrow} c_{I,\bar\alpha,\uparrow}+h.c),
	\label{jterm}
\end{equation}
where the mean-field order parameter $\lambda_{J_h}$ is defined as
$\lambda_{J_h}=-J_h\,\langle c^\dagger_{I,\alpha,\sigma} c_{I,\alpha,\bar\sigma} \rangle$.
Finally, if we insert Eq.~(\ref{jterm}) into the Hamiltonian $H$ and manually assign a real value to $\lambda_{J_h}$, we can confirm that phase \textbf{II}
with $\epsilon_\uparrow=-\epsilon_\downarrow$ is the preferred ground state for any small  $\lambda_J$.

\section{The Hamiltonian in momentum representation}
In this section, we Fourier transform $H^s$ into \textbf{k}-space for the 1-Fe per unit cell with the mean-field calculated order parameter $\lambda_{AOH}$ and the manually added perturbation terms $\lambda_0$ and $\lambda_1$.
For pedagogical reasons, we focus first on the spin polarized Hamiltonian for the spin component $\sigma$,
$H=\frac{1}{N}\sum_k \psi_{k}^\dagger\, \hat H\, \psi_{k}$,
where $\psi_{k}=(c_{k,1},c_{k,2})^T$ and $c_{1}/c_{2}$ stand for annihilating electrons on $d_{xz}$/$d_{yz}$ orbitals.
In  \textbf{k}-space the Hamiltonian  is conveniently written as
\begin{equation}
\hat H(\epsilon_\sigma)=E_0+\vec B\cdot\vec \tau,
\end{equation}
where $\vec B=(X,Y,Z)$ and $\vec\tau$ is the vector of Pauli matrices spanning the orbital $SU(2)$ space with
In the $2\times2$ matrix notation we can write explicitly,
\begin{equation}
\hat H(\epsilon_a)=
\left(
\begin{array}{cc}
 E_0+Z,	& X-iY\\
 X+iY,	& E_0-Z\\
\end{array}
\right).
	\label{Hless}
\end{equation}
Note that Eq.~(\ref{Hless}) has been used in the main text for the calculation of the Chern number.

The ancillary functions $E_0$, $X$, $Y$ and $Z$ are following from the manuscript,
\begin{equation}
\begin{aligned}
	E_0 =& 2 t_1 [ \cos(k_x)+\cos(k_y) ] + 2 t_6 [ \cos(2k_x)+\cos(2k_y) ] , \\
		&+ 2 (t_2+t_3) [ \cos(k_x) \cos(k_y) ]-\mu , \\
	X	=& 4 t_4 [ \cos(k_x) \cos(k_y) ]+ 2 (t_5+\delta t) [ \cos(k_x)+\cos(k_y) ] , \\
	Y	=& \epsilon\,(\,2 \lambda_{AOH} [\cos(k_x)-\cos(k_y)]+ \lambda_1), \\
	Z	=& 2 (t_2-t_3) [ \sin(k_x) \sin(k_y) ]+ \lambda_0.
	\label{element}
\end{aligned}
\end{equation}

%%%%%%%%%%%%%%%%%%%%%%%%%%
\begin{figure}
\includegraphics[scale=0.25,angle=0]{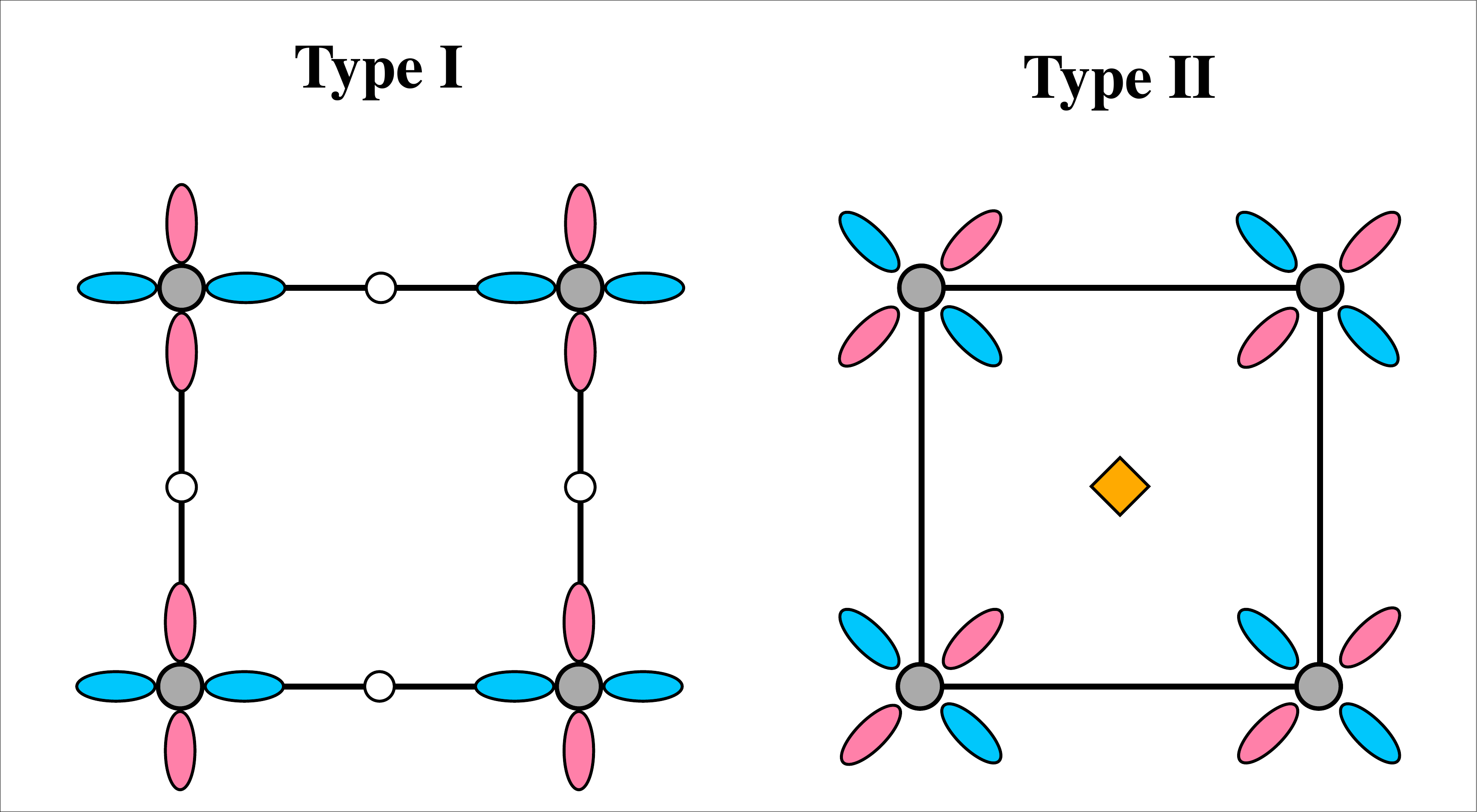}
\caption {(color online) {\bf Orbital order.}
Two generic types of orbital order in a lattice with C$_{4v}$ symmetry.
Type I  (left panel), the orbital orientation is along the NN bond direction with anions (open circles) between.
Type II (right panel), the orbital orientation is along the next-nearest neighbor (NNN) bond direction.
The blue (red) colored lobes represent the $d_{xz}$ ($d_{yz}$) orbitals of the Fe atoms (filled gray circles).
The filled yellow diamond indicates the anion atom, which prefers the Type II orbital locking in the Fe-only effective model.
}\label{pic:C4v}
\end{figure}

%%%%%%%%%%%%%%%%%%%%%%%%%
\begin{figure*}
\includegraphics[scale=0.45,angle=0]{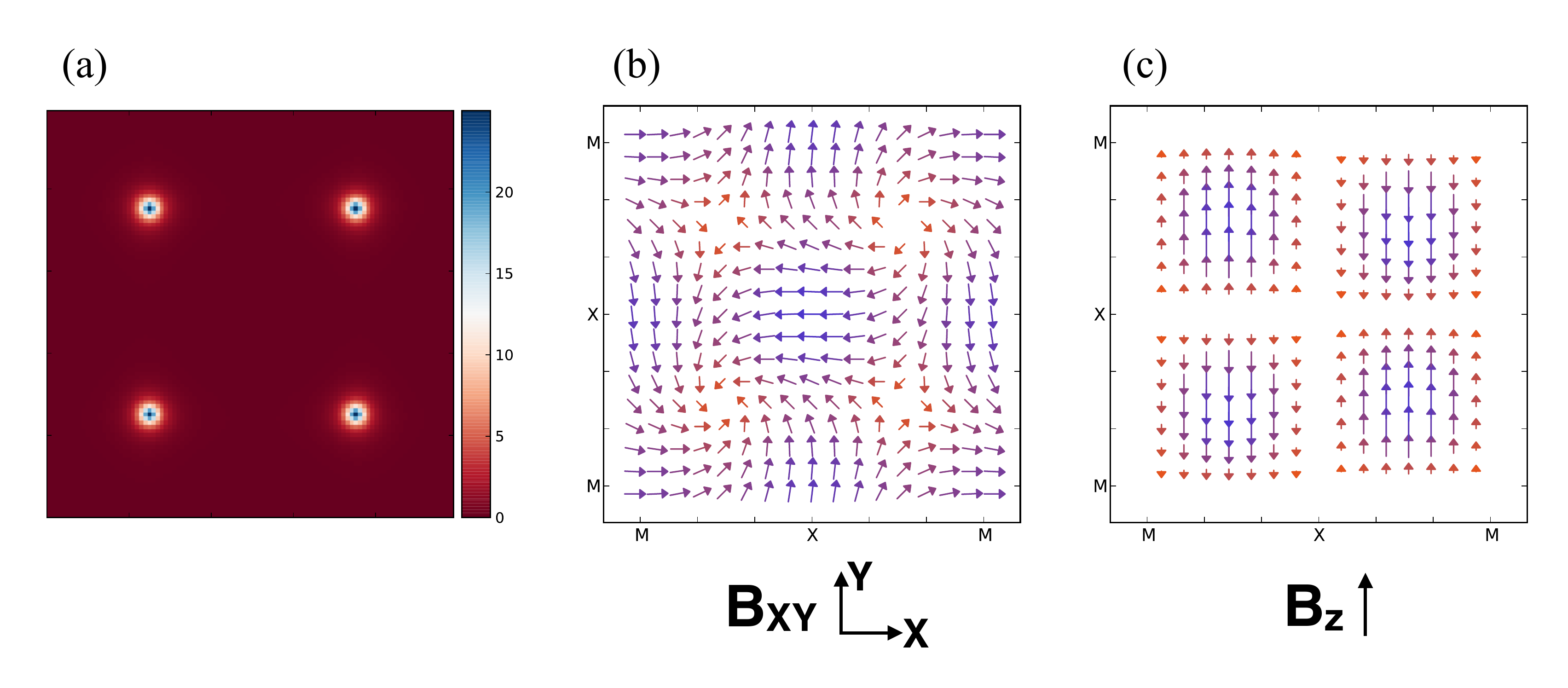}
\caption {(color online) {\bf Berry curvature and generalized magnetic field $\vec B$.}
(a) Color-map image of the Berry curvature of the simplified spinless Hamiltonian $\tilde H({\bf k})$ in the BZ.
(b) and (c) The components of  the complex auxiliary vector field $\vec B = (X, Y, Z)$  of  $\tilde H({\bf k})=\tilde E_0+\vec B\cdot\vec \tau$.
Each vector $B_{xy} = (B_x, B_y)$ with color-map red (small) to blue (large).
}\label{pic:BField}
\end{figure*}
%%%%%%%%%%%%%%%%%%%%%%%%%

We close this section by expanding the formulation of the spinless Hamiltonian to include spin degrees of freedom. The enlarged spin-orbital space becomes $SU(2)\times SU(2)$ or $\vec\tau \otimes \vec\sigma$.
In such a notation, an ``onsite Rashba"  or onsite spin-orbital coupling term enters on the off-diagonal entries of the 4$\times$4 matrix,
\begin{equation}
\hat H_s=
\left(
\begin{array}{cc}
 \hat H(\epsilon_\uparrow=1),	&  i\lambda_R\sigma_z\\
 -i\lambda_R\sigma_z,	& \hat H(\epsilon_\downarrow=-1)\\
\end{array}
\right).
	\label{Hful}
\end{equation}
Note that Eq.~(\ref{Hful}) has been used in the main text to calculate the stability of the mirror symmetry of the topological crystalline phase in two dimensions.

\section{Two types of orbital order locked in a lattice with C$_{4v}$ symmetry}
As shown in Fig.~\ref{pic:C4v} the lattice with
C$_{4v}$ symmetry
allows two different types of orbital order. The second type can generate Dirac cones along the $\Gamma$-$X$ direction in the bulk dispersion.
In what follows, we focus on the \textbf{k}-dependent inter-orbital hopping energies with C$_{4v}$ symmetry,
\begin{equation}
H^{xz,yz}=\sum_k [\epsilon_{xz}(k)\, c^\dagger_{xz,k} c_{xz,k} + \epsilon_{yz}(k)\, c^\dagger_{yz,k} c_{yz,k} +\cdots ,
\end{equation}
with
\begin{equation}
\begin{aligned}
	\epsilon_{xz}=-2\,t\,\cos({\bf k}\cdot \vec a_1),\\
	\epsilon_{yz}=-2\,t\,\cos({\bf k}\cdot \vec a_2),
\end{aligned}
\end{equation}
where $\vec a_{1,2}$ are orthogonal to each other and indicate the bond direction of the effective hopping term $t$.
If we choose $\vec a_1=(1,0)$ and $\vec a_2=(0,1)$, then this corresponds to type-I order with C$_{4v}$ symmetry.
On the other hand, for $\vec a_1=(1,1)$ and $\vec a_2=(-1,1)$ it becomes
\begin{equation}
	\hat H^{xz,yz}= -2\,t \big[ \cos(k_x)\cos(k_y) +\tau_z\sin(k_x)\sin(k_y) \big] ,
\end{equation}
which gives rise to the $\tau_z$ term in the spinless Hamiltonian in Eq.~(\ref{Hless}).

\section{Vortices as generators of the Berry flux and Chern numbers}
The anomalous orbital Hall effect of the spinless Hamiltonian with Chern number $\mathcal{C}=\pm 2$ is the combined effect of functions $X$, $Y$ and $Z$.
The Chern number can be calculated directly through the area integration of the Berry curvature\cite{Xiao2010}
\begin{equation}
	\Omega({\bf k})^\pm = i \frac{\langle \pm | \frac{\partial \hat H(k)}{\partial k_x} | \mp \rangle
							\langle \mp | \frac{\partial \hat H(k)}{\partial k_y} | \pm \rangle -
						(k_x \leftrightarrow k_y)}{(E_{\pm}-E_{\mp})^2}.
\end{equation}
In viewing the symmetry of $\vec B$, we know that the Dirac cones can be re-defined by taking any two components of ($X,Y,Z$).
To be specific, let us consider the simplified spinless Hamiltonian in Eq.~(2) of the manuscript with interaction renormalized coefficients,
\begin{equation}
\begin{aligned}
	\tilde E_0 =& 0 , \\
	\tilde X	=& t^x\big(4\,t^4\,[\,\cos(k_x)\,\cos(k_y)\,]+ 2\,t^5\,[\,\cos(k_x)+\cos(k_y)\,]\big) , \\
	\tilde Y	=& t^y\big(2\,\lambda_{AOH} [\cos(k_x)-\cos(k_y)]\big) , \\
	\tilde Z	=& t^z\big(2\,(t_2-t_3)[\,\sin(k_x)\,\sin(k_y)\,]\big) .
	\label{element2}
\end{aligned}
\end{equation}
Here $\tilde E_0$ can be regarded as an energy shift, because $E_0$ does not enter the wave function it does not contribute to the calculation of the topological invariant.
In the main text,  four Dirac cones are generated in the bulk bands
with renormalized hopping parameters $t^{x,y,z}=(1,0,1)$, where the band energy becomes
$E^{1,0,1}_\pm({\bf k})=\pm \sqrt{\tilde X^2+\tilde Z^2}$.
The four Dirac cones are located around the $M$ point of the BZ, satisfying the criterion $\tilde X^2+\tilde Z^2=0$.
{\it A key result of this work} is that a non-zero function $\tilde Y$ is a necessary, but not sufficient condition for a topological ground state. It can be viewed as a perturbation,
where the massless fermions acquire mass and a gap opens at the Dirac cones.
Here, if we choose the parameters $t^{x,y,z}=(1,1,0)$, then the dispersion becomes
$E^{1,1,0}_\pm({\bf k})=\pm \sqrt{\tilde X^2+\tilde Y^2}$, and the four Dirac cones are located in the BZ at
$(\pm \frac{\pi}{2},\pm \frac{\pi}{2})$. 
In addition, when we turn on a small $t^z$ or $\tilde Z$, then the degeneracy at the Dirac cones is lifted and the calculated Chern number becomes $\pm 2$ for each band.
The corresponding Berry curvature is shown as a color-map image in Fig.~\ref{pic:BField}(a), where four high intensity spots can be found at the positions of the Dirac cones.
If we regard these four Dirac cones as topological defects of a {\it generalized} magnetic field $\vec B$ in the Hamiltonian acting on the pseudo-spin degrees of freedom, that is,
${\vec B}\cdot \vec\tau$, where ${\vec B}=(\tilde X, \tilde Y, \tilde Z)$, then it is rather straightforward to map out the
$B_{xy}$ (in-plane) and $B_z$ (out-of-plane) components in Figs.~\ref{pic:BField}(b) and (c).
Knowing these two-dimensional (2D) vector maps, one can graphically solve for the Chern number by mapping the vector $\vec B$ around each singularity onto the Bloch sphere, following the procedure outlined by Bernevig \cite{Bernevig}.
Thus the $\vec B$-field winding around each topological defect (Dirac cone) contributes the winding number $2\pi \times \frac{1}{2} = \pi$, where the factor one-half stems from the spin $\frac{1}{2}$.
Hence the spinless Hamiltonian has the total Chern number
$\mathcal{C}=\pm (4 \times \pi) / 2\pi=\pm2$.

\section{Symmetry analysis of the time and mirror invariance}
In this section, we discuss the symmetry classification of phases \textbf{I} and \textbf{II} of the spinful Hamiltonian with spin degrees of freedom.
A detailed account of the symmetry operators used in the main text is given.
The spinful Hamiltonian in  \textbf{k}-space of fermions is defined by
$H^{s}=\phi_{\bf k}^\dagger \hat H_{s}({\bf k}) \phi_{\bf k}$, where $\phi_{\bf k}=(c_{{\bf k},1\uparrow}, c_{{\bf k},2\uparrow}, c_{{\bf k},1\downarrow}, c_{{\bf k},2\downarrow})^T$.
We re-write $\hat H_s$ as a direct tensor product of Pauli matrices in the combined orbital pseudo-spin and spin spaces,
$\hat H^{I,II}_s = X\,\tau_x\otimes I+Z\,\tau_z\otimes I+\hat H^{I,II}_{AOH}$,
where the orbital flux term, $\hat H^{I,II}_{AOH}$, of phases
\textbf{I} and \textbf{II} is either
$\hat H^{I}_{AOH}=Y\tau_y\otimes I$
or
$\hat H^{II}_{AOH}=Y\tau_y\otimes\sigma_z$. Here $I$ is the $2\times 2$ unity matrix in spin space.

\subsection{Intrinsic inversion symmetry and TR symmetry violation of spinless $\hat H$}
We start our symmetry analysis by noting that the quasi-2D Hamiltonian of spinless fermions, $\hat H$, in a tetragonal system has intrinsic inversion symmetry
$\hat H ({\bf k})=\hat H (-{\bf k})$. This corresponds to a $180^o$ rotation in the $k_x$-$k_y$ plane.
Moreover, for the spinless Hamiltonian the time-reversal (TR) operator is given by the charge conjugation operator, $\mT = \mK$, and satisfies the relation
\begin{equation}
\begin{aligned}
\mT\,\hat H[\lambda_{AOH}]({\bf k})\,\mT^{-1}	=& \hat H[-\lambda_{AOH}](-{\bf k})\\
												=& \hat H[-\lambda_{AOH}]({\bf k}),
\end{aligned}
\end{equation}
which tells us that the TR symmetry is violated, because it reverses the orbital current direction from $\lambda_{AOH} \to  -\lambda_{AOH}$.
Hence it is not too surprising that the TI is quite robust against perturbations $\lambda_0$ and $\lambda_1$, which break TR symmetry, too.

\subsection{Reflection invariance of spinless $\hat H$}
In addition to the inversion symmetry, the spinless quasi-2D Hamiltonian is invariant under reflections. The two reflection axes $x$ and $y$  obey the parity operation
\begin{equation}
\mP\,\hat H(k_x,k_y)\,\mP^{-1} = \hat H(\pm k_x,\mp k_y),
\end{equation}
respectively, with $\mP=\tau_x \mK$.
This statement is universally true for our model Hamiltonians and applies also to phases \textbf{I} and \textbf{II} of the spinful Hamiltonian.
Note that in 2D the parity operation is a reflection which only acts on orbital degrees of freedom.

\subsection{Parity and mirror invariance of phase \textbf{II} of spinful $\hat H_s$}
The spinful Hamiltonian for fermions with spin degrees of freedom satisfies mirror symmetry operations in the enlarged orbital $\times$ spin space. Since the reflection and TR operators flip the spin of the fermion, they must be defined in the enlarged $SU(2)\otimes SU(2)$ space as
$\mM=\mP \otimes \mT$ and $\mT=-i\sigma_y\mK$, where $\mP$ ($\mT$) is the operation with respect to the {orbital} ({spin}) degree of freedom.

To summarize the key results of our symmetry analysis, our spinful model Hamiltonian, $H=\sum_k \phi^\dagger_k \hat H^{II}_s(k) \phi_k$, of phase \textbf{II} is invariant under the TR operation,
\begin{eqnarray}
&\mT\,\hat H^{II}_s({\bf k})\,\mT^{-1} = \hat H^{II}_s(-{\bf k}) = \hat H^{II}_s({\bf k}),
\end{eqnarray}
and the mirror operations, 
\begin{eqnarray}
&\mM\,\hat H^{II}_s(k_x,k_y)\,\mM^{-1} = \hat H^{II}_s(\pm k_x,\mp k_y) ,
\end{eqnarray}
connecting all four quadrants of the BZ.

\subsection{Even and odd parity subspaces of phase \textbf{II} of spinful $\hat H_s$}
For the spinful Hamiltonian, the Chern number is only meaningful for phase \textbf{I}. This can be seen from its non-zero Chern number
$\mathcal{C}[\hat H^{I}_s]=-\mathcal{C}[\mM\,\hat H^{I}_s\,\mM^{-1}]=-\mathcal{C}[\mT\,\hat H^{I}_s\,\mT^{-1}]=\pm 4$.
Consequently, $\hat H^{I}_s$ has two distinguishable degenerated states of $\mathcal{C}=\pm 4$,
which can be mapped onto each other.

On the other side, phase \textbf{II} also has two distinguishable degenerated states, however, these two states cannot be distinguished by the Chern number, because $\mathcal{C}[\hat H^{II}_s]=0$.
Thus, we need to further examine its symmetry properties to see whether it is topological  or not. A very direct and useful check is to see whether the system has a $Z_2$-like invariant index. This symmetry has been widely used in the search for 2D and 3D topological insulators, because there exist general methods to calculate the $Z_2$ topological invariant, especially when the Hamiltonian exhibits inversion symmetry.\cite{Fu2007a, Moore2007, Roy2009, Fu2007b}
In phase \textbf{II}, a close inspection of $\vec B$ shows that the operator $\mM$ commutes with the orbital-flux term $\hat H^{II}_{AOH}$ and $\hat H^{II}_s$.
The {even} parity subspace is described only by contours along the boundaries of the quadrants of the BZ, for example, 
$\mathbb{C} = \{ \Gamma \to X \to M \to \bar X \to \Gamma \}$, and satisfies the commutation relation for any ${\bf k} \in \mathbb{C}$ with
$\mM\,\hat H^{II}_s(\mathbb{C}) \,\mM^{-1} = \hat H^{II}_s(\mathbb{C})$.
The {odd} parity subspace is located only at high-symmetry points given by the roots of the Pfaffian. For example, for the parameters chosen for the noninteracting case and setting the term $E_0=0$ (because the topological property is manifested only in $\vec B$), the four points in the set
$\Lambda_n = \{ (\pm\frac{\pi}{2},\pm\frac{\pi}{2})  \}$
satisfy each the anti-commutation relation, $\mM\,\hat H^{II}_s(\Lambda_n)\,\mM^{-1} =-\hat H^{II}_s(\Lambda_n)$.
Anywhere else in the BZ the Hamiltonian has a mixture of odd and even subspace terms.
It is precisely this $\pm$-parity symmetry that motivated us to construct the $Z_2$-like topological invariant in the main text in order to test and quantify the topological ground state with vanishing Chern number.


\begin{thebibliography}{10}
\bibitem{footnote2} The original model of~\cite{YTai} assumed the 2-Fe per unit cell Hamiltonian. However, a unitary transformation maps the 2-Fe onto the 1-Fe per unit cell model Hamiltonian in the absence of the spin density wave~\cite{HChen}.
\bibitem{footnote3} One might observe that the Dirac points sink below the Fermi surface and do not cause topological behavior. This is not true for our case,
since the real part of the order parameter from the the mean-field calculated NN-Coulomb interaction can lift the Dirac point toward the Fermi surface.
This results in a direct gap when the imaginary part is non-zero, see the SM for more detail.
\end{thebibliography}
\end{document}